\DocumentMetadata{
  pdfversion=2.0,pdfstandard=ua-2,
  testphase={phase-III,firstaid,math,title}
}

\documentclass[sigconf,screen]{acmart-tagged}
\AtBeginDocument{%
  }

\usepackage{microtype}
\usepackage[font={small,bf}]{caption}

\begin{document}

\title[How Neurotypical \& Autistic Children Interact Nonverbally with Anthropomorphic Agents in Open-Ended ...]{How Neurotypical and Autistic Children Interact Nonverbally with Anthropomorphic Agents in Open-Ended Tasks}







\author{Chuxuan Zhang}
\email{chuxuan\_zhang@sfu.ca}

\author{Bermet Burkanova}
\email{bermet\_burkanova@sfu.ca}

\affiliation{%
  \institution{Simon Fraser University}
  \city{Burnaby}
  \country{Canada}
}

\author{Lawrence H. Kim}
\email{lawkim@sfu.ca}

\author{Grace Iarocci}
\email{giarocci@sfu.ca}

\affiliation{%
  \institution{Simon Fraser University}
  \city{Burnaby}
  \country{Canada}
}

\author{Elina Birmingham}
\email{elina\_birmingham@sfu.ca}

\author{Angelica Lim}
\email{angelica@sfu.ca}

\affiliation{%
  \institution{Simon Fraser University}
  \city{Burnaby}
  \country{Canada}
}





\begin{abstract}
What nonverbal behaviors should a robot respond to? Understanding how children—both neurotypical and autistic—engage with embodied artificial agents is critical for developing inclusive and socially interactive systems. In this paper, we study ``open-ended" unconstrained interactions with embodied agents, where little is known about how children behave nonverbally when given few instructions. We conducted a Wizard-of-Oz study in which children were invited to interact nonverbally with 6 different embodied virtual characters displayed on a television screen. We collected 563 (141 unique) nonverbal behaviors produced by children and compare the children’s interaction patterns with those previously reported in an adult study. We also report the presence of repetitive face and hand movements, which should be considered in the development of nonverbally interactive artificial agents.

\end{abstract}

\ccsdesc[500]{Human-centered computing~Empirical studies in HCI}

\keywords{nonverbal interaction, social robotics, children robot interaction}

\setcopyright{cc}
\setcctype{by}
\acmDOI{10.1145/3776734.3794359}
\acmYear{2026}
\copyrightyear{2026}
\acmISBN{979-8-4007-2321-6/2026/03}
\acmConference[HRI Companion '26]{Companion Proceedings of the 21st ACM/IEEE International Conference on Human-Robot Interaction}{March 16--19, 2026}{Edinburgh, Scotland, UK}
\acmBooktitle{Companion Proceedings of the 21st ACM/IEEE International Conference on Human-Robot Interaction (HRI Companion '26), March 16--19, 2026, Edinburgh, Scotland, UK}
\received{2025-12-08}
\received[accepted]{2026-01-12}


\maketitle

\section{Introduction}
\begin{figure}[h]
    \centering
    \captionsetup{font=small}
    \includegraphics[width = 0.42\textwidth, alt={The left figure shows the room setup for the study. A yellow line was taped on the ground to indicate interaction area, and a smart TV is placed against wall which display the interactive virtual characters. A laptop was placed under the TV screen which provide the TV with image source; The right figure shows the six characters used in the study: a human, a robot, a penguin, a fish, a toilet and a banana}]{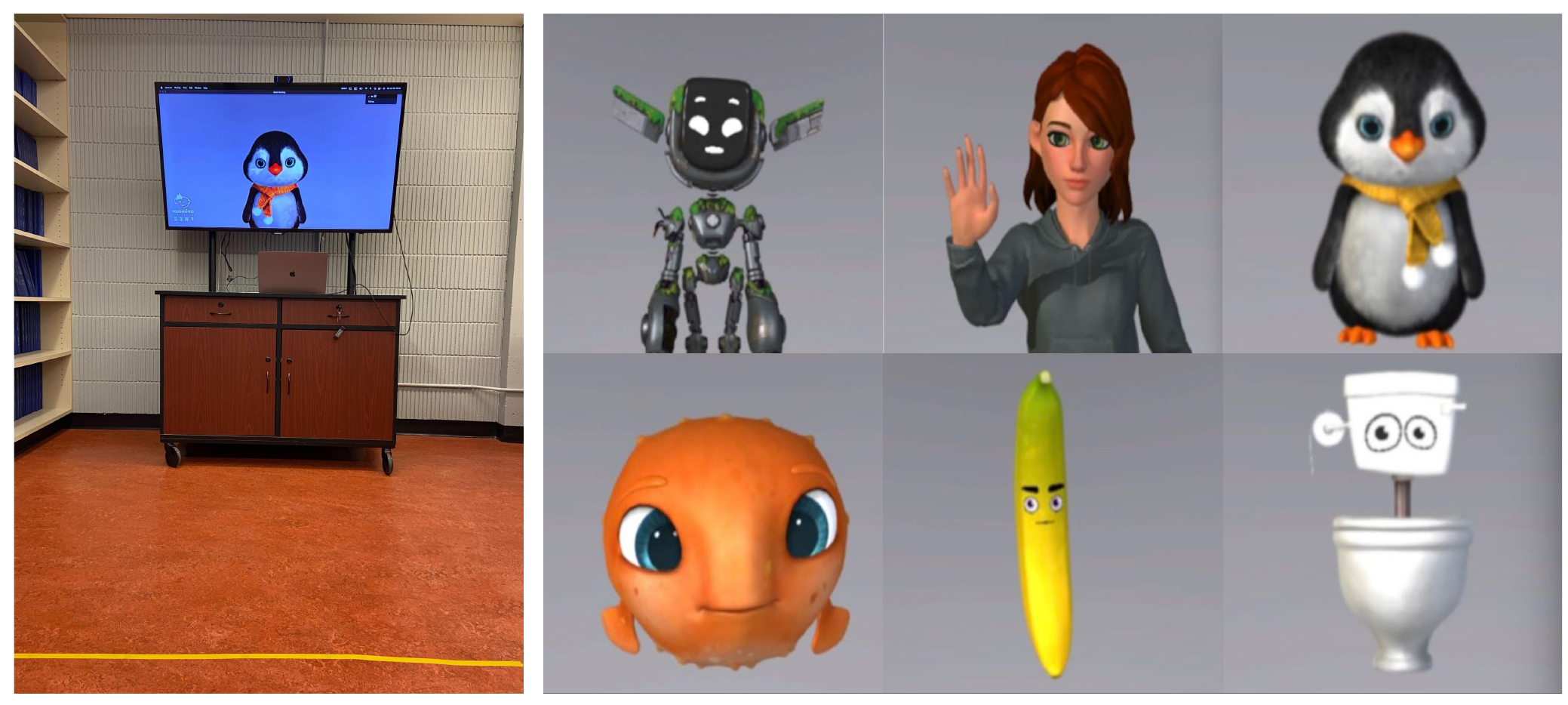}
    \caption{\textbf{Left: Physical setup of the study room. Right: 6 virtual characters used in the study.}}
    \label{fig:setup}    
\end{figure}

Imagine a child waving to a robot. A social robot may be programmed to respond to this gesture, but what other nonverbal behaviors do children make that a robot should respond to? Embodied artificial agents, ranging from humanoid robots to animated virtual characters, are increasingly present in public spaces for entertainment \cite{kudoh2008entertainment, xiaochun2025research} and service applications \cite{song2022role, moriuchi2024role}. Prior work has examined how adults nonverbally test the physical, emotional, and social capacities of agents with varied morphologies \cite{zhang2024react}, but these findings may not translate directly to children’s behaviors.


To support embodied agents that can understand a wide range of children's behaviors, we conducted a study with children aged 7–12, including both neurotypical and autistic participants. Building on prior adult-focused work \cite{zhang2024react}, we analyzed how children initiate and sustain nonverbal engagement with six virtual characters, revealing both shared and distinct behavioral patterns across age and neurodevelopmental groups. This paper contributes to child-centered interaction design by:

\begin{enumerate}
    \item identifying similarities and differences in nonverbal interaction across age and neurodevelopmental profiles;
    \item providing a set of child-generated nonverbal behaviors relevant for gesture and activity recognition;
    \item offering insights for designing inclusive, socially adaptive embodied systems that can better interpret and respond to diverse children’s behaviors.
\end{enumerate}

\section{Related Work}
\textit{Human Nonverbal Interactive Behaviors}. Nonverbal behavior plays a central role in how humans perceive and interact with embodied agents. Prior work shows that people are more likely to engage with robots capable of producing expressive and meaningful nonverbal signals \cite{han2012investigating}. In order to generate such behaviors, the intelligent agents should also perceive and understand the nonverbal actions that humans perform during interaction~\cite{10.5555/1558109.1558314}\cite{10.5555/2615731.2617415} \cite{yalccin2020m}.
Several datasets have attempted to capture naturalistic human–robot interactions within task-orientated contexts (e.g., conversation, tutoring), including UE-HRI \cite{10.1145/3136755.3136814} and PInSoRo \cite{lemaignan2018pinsoro}, providing valuable annotated corpora of social interactions. Complementing these efforts, Zhang et al. \cite{zhang2024react} offered a data-driven characterization of the nonverbal “testing” behaviors that adults use when probing the agents' interaction awareness in a task-free environment.
However, their study focused exclusively on adults. Our work directly extends this line of inquiry by examining a broader and younger population—specifically school-aged, neurotypical and autistic children—to deepen our understanding of how diverse user groups engage nonverbally with embodied systems.


\textit{Children-Robot Interaction}. Child–robot interaction (CRI) has long been a focal point in domains such as education \cite{jones2015empathic,vogt2017child,jacq2016building} and entertainment \cite{gordon2015designing, aaltonen2017hello}. Children are widely recognized as a distinct user group in HRI: as Belpaeme et al. note, ``children are not just small adults” \cite{belpaeme2013child}. Their developmental stage, cognitive abilities, and cultural exposure shape their expectations and interaction styles with artificial agents. Compared to adults, children tend to anthropomorphize robots more readily \cite{turkle2006encounters}, often treating them as living creatures. 
While children and adults differ socially and physically, comparative research on their non-verbal interactions is limited; we fill this gap by testing children with a previously adult-validated agent set.


\textit{Autism and Robotics}. Autism spectrum disorder (ASD) is characterized by differences or challenges in social skills \cite{williams2007social, holloway2014social}, including nonverbal behaviors such as gesture use, eye contact, and affective expression \cite{hodges2020autism, syu2020relationship, tomczak2018support, wiklund2016interactional}. These characteristics have motivated the growth of robot-assisted autism therapy (RAAT) to support social skill development \cite{alabdulkareem2022systematic,rakhymbayeva2021long}.
Autistic children often show strong engagement with robots \cite{bauminger2003peer}. For instance, Barnes et al. \cite{barnes2021child} found that autistic children participated more actively in a musical dance game with a robot than typically developing peers. A case study reported that autistic children initiate interactions more frequently and perform more complete movements when engaging with robots than with human partners \cite{aryania2020social}. Lytridis et al. \cite{lytridis2020robot} similarly observed the eagerness to participate and sustained engagement among autistic children when interacting with multiple robots. 
While these findings demonstrate the potential of robots as social partners for autistic children, most prior work examines task-oriented or intervention-driven scenarios. In contrast, our study focuses on spontaneous, task-free interaction.

\section{Methodology}
Children demonstrate social interaction abilities that differ from those of adults due to differences in neural \cite{dosch2010learning} and cognitive \cite{carpendale2004constructing} development. To enable a parallel comparison with the adult study \cite{zhang2024react}, we used the same set of virtual characters with 2 humanoids, 2 animals and 2 objects (Fig.~\ref{fig:setup}), but adapted the study question, protocol, and instructions to ensure comprehension for children. To investigate similarities and differences in interactive behaviors across demographic groups, we posed the following \textbf{research question}:
\emph{How do neurotypical and autistic children play or interact with a virtual agent?}



\subsection{Study Design}
We conducted a Wizard-of-Oz study with 14 participants. The study was approved by the institution’s research ethics board.

\subsubsection{Participants. }
Participants were recruited through a university-run Social Science Camp designed for children aged 7–12, with or without a diagnosis of ASD. Prior to the camp, written consent was obtained from each child's guardian. Before the study session, we also obtained verbal assent from the participants and informed them that they could withdraw at any time without penalty.
We collected data from 14 children (Sex: 8/6 male/female; Age: $9.36 \pm 1.39$; Developmental Diagnosis of w/wo ASD: 10/4). 


\subsubsection{Study Prompt. }
To ensure question comprehension across ages, we adapted the instructional wording. For younger children (age $<$ 10), we used: “What would you do to play with this character?” For older children, we used: “What would you do to interact with this character?” Based on prior camp experience, we used play-oriented language for younger participants and adult-like phrasing for older children, ensuring age-appropriate engagement without altering the experimental goal.

\subsubsection{Study Procedure. }
To enable direct comparison with the adult dataset, participants were instructed to interact with each virtual character for approximately one minute. In practice, compliance varied: some sessions were extended due to strong interest in the characters, while others ended early due to reduced enthusiasm. The researcher timed each session and announced its end, but participants determined the actual duration of their interaction.
One participant (P7) primarily used drawing and physical objects rather than facial or bodily gestures. Their data were therefore excluded from the gesture analysis and are discussed separately. The final dataset includes nonverbal gesture data from 13 participants.



\subsubsection{Software. }

We used the same technical setup as the adult study for teleoperation with a single Wizard~\cite{zhang2024react}. Three off-the-shelf software tools were used: \emph{Animaze}\footnote{https://www.animaze.us/} for character animation and facial motion mapping, \emph{Webcam Motion Capture}\footnote{https://webcammotioncapture.info/} for tracking upper-body movement (fingers, arms, torso), and \emph{Zoom} \footnote{https://zoom.us/}for real-time visual interaction between participant and teleoperator.

\subsection{Data Collection and Analysis}
All interaction sessions were video-recorded, and we applied the same segmentation, annotation and thematic analysis workflow as~\cite{zhang2024react}, with 2 raters and discussion until agreement, and a 3rd rater (teleoperator) as a tie-breaker. During some sessions, participants asked what they could do with the virtual character; to maintain engagement, the researcher occasionally offered simple prompts (e.g., “you could try to pet it”). Accordingly, we added an additional annotation label—“instructed (yes/no)”—to mark whether a behavior was produced following a researcher's suggestion.

\section{Result and Discussion}
\begin{figure}[!t]
    \centering
    \captionsetup{font=small}
    \includegraphics[width = 0.40\textwidth]{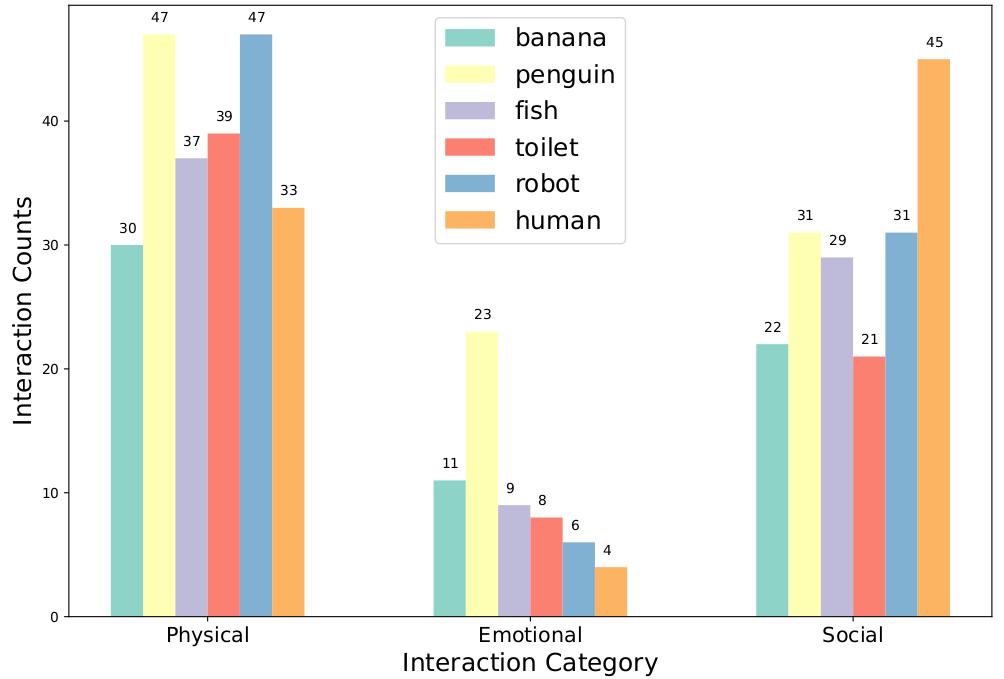}
    \caption{\textbf{Interaction behavior counts categorized the type of interaction.}}
    \label{fig: minor behavior count}
\end{figure}

In this section, we address the research question \textbf{"How do autistic and neurotypical children play or interact with a virtual agent?"} by comparing children’s behaviors with those reported in the adult study \cite{zhang2024react}, and then highlight the additional observations unique to this work. In total, 563 interactions were observed during the study (Fig. \ref{fig: minor behavior count}), including 537 interactions initiated by the participants and 26 interactions initiated by the agents. From the 537 interactions started by the participants, there are 233 character-agnostic physical instances (89 unique), 63 character-agnostic emotional instances (18 unique), 177 character-agnostic social instances (34 unique), and 64 character-specific behaviors.

\begin{table*}[!h]
\small
\centering
\caption{\textbf{Physical behaviors initiated by minors; behaviors absent in the adult study in bold; counts in parentheses.}}
\captionsetup{font=small}
\begin{tabular}{p{3.5cm}|p{13.3cm}}
    \hline
    \textbf{Category} & \textbf{Nonverbal Physical Behaviors} \\
    \hline
        posture -- full body (47)& tilt/lean body to the side (17); swing/wiggle (9); turn around (5); bend body forward (3); rotate/turn/swing the upper body (3); jumping jack (3); bend body to the side (2); lean body forward/backward (1); squat (1); turn body to the side (1); \textbf{bend torso forward and downward till hands and head on the ground} (1); \textbf{plank} (1)\\ 
    \hline
    
    posture -- head/face (83) & stick out tongue (18); open mouth (14); pout mouth (12); shake head (6); tilt head (5); blink eye(s) (4); squint/squeeze eyes (4);  bare teeth (3); \textbf{eat} (2); open and close mouth (2); \textbf{make face} (1); raise one eyebrow (1); close one eye (1); \textbf{shout} (1); \textbf{look back} (1); \textbf{bend neck forward and backward} (1);  \textbf{bubble} (1); wink (1); frown (1); \textbf{shake hair} (1); \textbf{bite} (1); \textbf{open mouth in different shapes} (1); \textbf{move the tongue up and down} (1)\\ 
    \hline
    
    posture --  arm (28)  & cross arms (5); open arms (5); raise arms (3); arms drawing circles (3); wave arms (2); \textbf{randomly move arms} (2); swing arms (2); forearms rotate (2); shake arms (2); arm raised pointing to the side (1); \textbf{arm dive} (1)\\ 
    \hline
    
    posture -- hand (58) & flap hands (9); \textbf{look through ok gesture hand} (7); hands/fists rub other body parts (6); clap (5); wave hand (5); hands on other parts (4); hands touch other body parts (5);\textbf{make fists} (2); raise hand(s) (2); fingers drawing circles (2); \textbf{wrap the fist with another hand} (1); \textbf{hands hover over eyes} (1); hand shakes (1); show fingers (1); \textbf{point to the character quickly} (1); \textbf{fingers pinch cheeks (1)}; \textbf{clap hands in front of and behind the body} (1); hand(wrist) rotate in a small circle (1); \textbf{fists hold in front of chest (cat like gesture)} (1); groom hair (1); scratch body parts (1)\\ 

    \hline
    posture -- lower body (15)& \textbf{foot fire} (3); lift leg(s) up (2); squat (2); side kick (1); leg swing (1); shake legs (1); lift leg up to the side (1); knee on the floor (1); bend knees (1); lift one leg with hands (1); stand on toes (1)\\ 
    
    \hline  
    proxemics (27) &  jump (16); step to the side (4); walk to the left and right (3); step forward/backward (2); run (1); \textbf{step forward and backward while squatting} (1)\\ 
    
    \hline
    physical contact (16) &  grab (6); poke (5); \textbf{tickle} (3); squeeze (1); \textbf{press} (1)\\
    \hline      

\end{tabular}

\label{tab: minor physical}

\end{table*}

\begin{table}[!h]
\small
\captionsetup{font=small}
\centering
\caption{\textbf{Emotional behaviors initiated by minors; behaviors absent in the adult study in bold; counts in parentheses.}}
\begin{tabular}{p{2cm}|p{5.7cm}}
    \hline
    \textbf{Category} & \textbf{Nonverbal Emotional Behaviors} \\
    \hline
    affectionate (38) & pet (28); caress (2); heart gesture (5); kissing mouth (1); hug (2)\\
    \hline
    angry (19) & frown (8); pout mouth (3); \textbf{shout} (2); side eyes (2); \textbf{bare teeth} (1); \textbf{hands on hips} (1); \textbf{bite lips} (1); stare (1)\\
    \hline
    happy (8) & smile (7); laugh (1)\\
    \hline
    sad (5) & pout mouth (3); sad face (1); \textbf{squint eyes} (1)\\
    \hline

\end{tabular}

\label{tab: minor_emotional}
\end{table} 
\begin{table}[!h]
\small
\centering
\caption{\textbf{Social behaviors initiated by minors; behaviors absent in the adult study in bold; counts in parentheses.}}
\begin{tabular}{p{3cm}|p{4.7cm}}
    \hline
    \textbf{Category} & \textbf{Nonverbal Social Behaviors} \\
    \hline
    greetings (63) & wave hands (55); hand shake (8)\\
    \hline
    attack (34) & punch (13); kick (8); \textbf{poke} (4); slap (8); \textbf{grab and drag} (1)\\
    \hline
    insult/ taunt (20) & \textbf{stick tongue out} (7); \textbf{loser sign in front of the forehead} (6); \textbf{crooked mouth, side eyes} (5); \textbf{make face} (1); \textbf{side eyes(1)}\\
    \hline
    you (17) & point to the character (17)\\ 
    \hline
    dance (12) & casual dance (11); dab (1)\\   
    \hline
    sleep (8) & palms together, head rests on hands, sleep (8)\\    
    \hline
    deictic (7) & point to some direction (7)\\
    \hline
    I/me (6) & point to themselves(5); hand on chest(1)\\
    \hline
    approval (5) & nod (5)\\   
    \hline    
    cheering (4) & clap (2); high five (2)\\
    \hline
    instructional (3) & perform certain action and point to the character to instruct the character to replicate the same action (2); \textbf{slowly demonstrate the completion of a sequence of movement}(1)\\    
    \hline
    \textbf{to surprise} (3)	& shout(1); suddenly raise hands up and put them down (2)\\    
    \hline
    ok (2) & ok gesture (2)\\
    \hline
    denial (1) & shake head (1)\\
    \hline
    draw attention (1) & finger snap (1)\\
    \hline
    entertaining (1) & make face (1)\\
    \hline
    good job (1) & thumbs up (1)\\
    \hline
    I don't know (1) & shrug(1)\\
    \hline
    listen (1) & putting hand close to the ear (1)\\
    \hline
    come closer (1)	& pull fingers to themselves (1)\\    
    \hline
        
\end{tabular}
\label{tab: minor_social}
\end{table}

\textbf{\textit{Physical, Emotional, and Social Behaviors}} In total, the children produced 89 distinct physical behaviors, 18 emotional behaviors, and 34 social behaviors. Compared with the adult dataset, children displayed a smaller overall variety of nonverbal behaviors. However, we identified several behaviors that were not present in the adult study (Tables~\ref{tab: minor physical}, \ref{tab: minor_emotional}, \ref{tab: minor_social}). For example, children frequently engaged in playful and imaginative actions—such as exaggerated faces or lying on the floor—indicating a more exploratory interaction style. No significant group difference (neurotypical vs. autistic) was observed in behavior counts.

We conducted two-tailed $\chi ^2$ test on the behavioral frequency count among all 6 characters, and found a significant difference $\chi ^2 \left(10 \right)= 26.92$, $p = 0.0026$.
We observed a significantly large positive adjusted residual ($adjusted\_residual = 3.49$, $p = 0.0084$) in \emph{social} behaviors from the human condition, and a large positive adjusted residual ($adjusted\_residual = 3.34$, $p = 0.0151$) in \emph{emotional} behaviors from the penguin condition after Bonferroni adjustment.


Emotional behavior differences across characters were largely driven by the high frequency of petting penguin action. This pattern did not appear in the adult study, making it a child-specific repeated emotional behavior.

\textbf{\textit{Drawing and Interacting with the environment}}
Søndergaard and Reventlow suggested, ``With their drawings, the children were able to express feelings, sentiments, and experiences that were difficult to articulate in words but not equally difficult to recall as a physical and mental experience or to draw on paper." \cite{sondergaard2019drawing}. This was evident in the behavior of Participant 7 (P7), who interpreted the instruction to avoid sound or language as meaning that the virtual characters could ``see" visual content. Despite encouragement to use facial and body gestures, P7 primarily communicated through drawing and by interacting with objects in the room.

\begin{figure}[!t]
    \captionsetup{font=small}
    \centering
    \includegraphics[width=0.8\linewidth, alt={The figure shows what a participant drew during the interaction. From left to right: a igloo with a penguin in it; a blue ice pile; a green ghost and a brown toilet with the feces in it; a smiling yellow banana}]{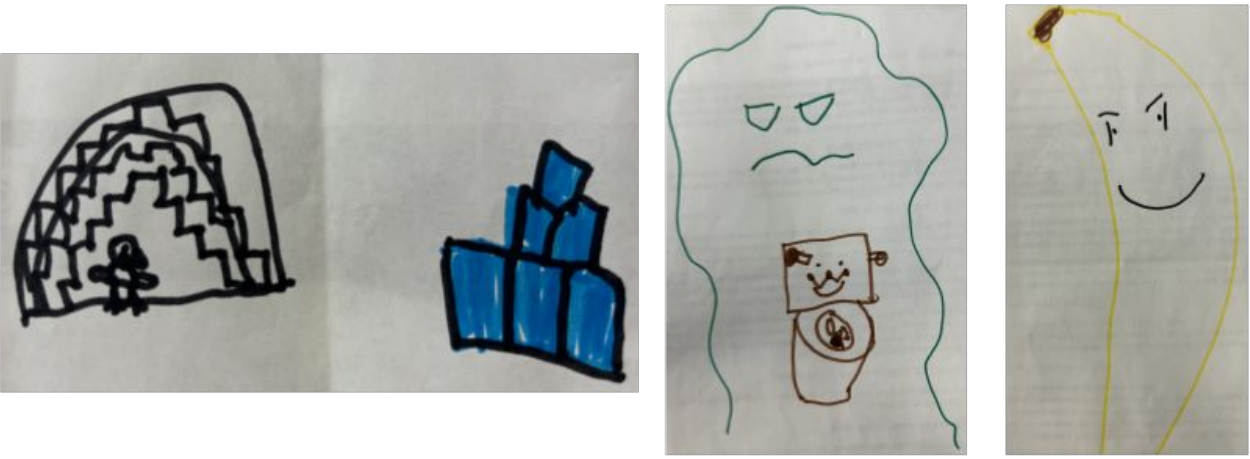}
    \caption{\textbf{Drawings created by participant 7 correspondingly during the penguin, the toilet, and the banana interactions.}}
    \label{fig: drawing}
\end{figure}


During the penguin session, P7 drew an igloo and a small penguin, intending to ask the virtual penguin about its family. To further clarify the context, they added a pile of ice cubes beside the igloo. Similar strategies were used with other characters: they showed the banana character a drawn banana, and they created a drawing depicting excrement to ask the toilet character whether it ``contained poo".

Beyond drawing, P7 also experimented with physical objects and the room environment. When the banana character did not respond strongly to the drawing, P7 presented a real banana from their lunchbox to the screen. They also took a photo of the on-screen banana and showed it back to the character. While interacting with the toilet, they turned off the lights in the study room and displayed a hand-drawn ghost to the character as a playful ``prank". Additionally, they offered their snack bar to the fish character.


While adults occasionally used objects like phones, only children engaged in drawing, environmental manipulation, and creative prop use during study. These behaviors, though outside the original task, suggest that HRI/HCI systems should accommodate creative, multimodal, and environmentally embedded interaction strategies.


\begin{figure}
    \centering
    \captionsetup{font=small}
    \includegraphics[width=0.75\linewidth, alt={The figure shows the 4 repetitive behaviors observed in the study. They are flicking, blinking, poking and swiping.}]{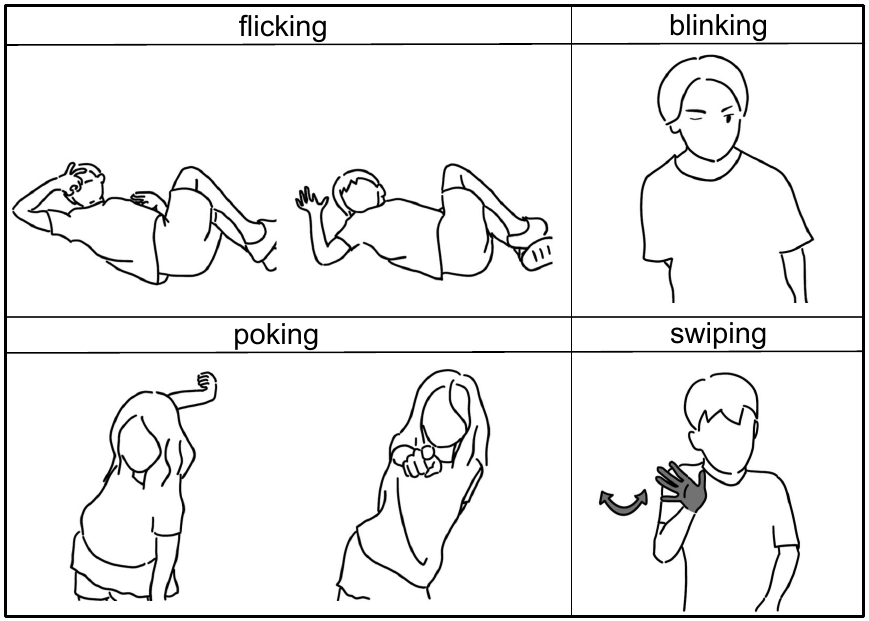}
    \caption{\textbf{Repetitive behaviors observed during study.}}
    \label{fig: repetitive}

\end{figure}

\textbf{\emph{Repetitive Behavior}}  Repetitive behavior is one of the core diagnostic features of ASD \cite{10665-42980, american1994diagnostic, turner1999annotation}. During annotation, such behaviors stood out due to their high frequency and ambiguous intent. In total, 4 participants - 1 neurotypical and 3 autistic—displayed repetitive actions (Fig.~\ref{fig: repetitive}).

Participant P11 (neurotypical) repeatedly formed a thumb–index circle, lifted it to one eye as if looking through a lens, then flicked the fingers outward while swinging the forearm away. This sequence resembled a “telescope” gesture, suggesting exploratory or social intent, but its continued repetition without waiting for a response left the meaning unclear.

Among the autistic group, 3 distinct sensorimotor behaviors were noted. P2 repeatedly pointed toward the character with force, resembling a firm poking motion. P5 showed rapid right-hand flapping throughout the sessions. Although the motion resembled a common autistic motor mannerism, its interpretation remains ambiguous. Our researcher with expertise in autism suggested that the participant might have been using this movement to test whether the agent would respond, while a camp organizer noted that the participant sometimes held their hands together during the interaction, implying that this could also be a mitigating behavior used to suppress the flapping. P8 displayed frequent, forceful blinking behaviors. Compared with P11’s complex sequence, these 3 behaviors were brief and mechanically simple, but their meaning is unclear and may reflect either motor mannerisms or interactive responses depending on context.

These observations underscore the need for interactive agents to distinguish communicative behaviors from non-interactive repetitive movements, ensuring contextually appropriate and supportive responses.



\section{Conclusion}
In this study, we identified a set of 141 unique nonverbal behaviors that autistic and neurotypical children use to try to interact with anthropomorphic agents. These preliminary results include some not found in the adult study (e.g. sticking tongue out, loser sign, etc.). However, the small sample size (n=14) and uneven neurotypical–autistic distribution limits the ability to compute differences between the two groups. Future work will include a larger and more balanced participant pool, as well as studies with physical robots and richer environments to better understand what spontaneous child behaviors robots and agents should respond to.



\bibliographystyle{ACM-Reference-Format}
\bibliography{references, chi}

\appendix

\end{document}